\documentclass[12pt,preprint]{aastex}

\newcommand{\figref}[1]{Figure~#1}

\usepackage[]{natbib}

\slugcomment{ ApjL}

\shorttitle{Solar Wind Drag and the Kinematics of ICMEs}
\shortauthors{Maloney et al.}

\begin{document}


\title{Solar Wind Drag and the Kinematics of Interplanetary Coronal Mass Ejections}


\author{Shane A. Maloney and Peter T.  Gallagher}
\affil{School of Physics, Trinity College Dublin, College Green, Dublin 2, Ireland}
\email{maloneys@tcd.ie}




\begin{abstract}
Coronal mass ejections (CMEs) are large-scale ejections of plasma and magnetic field from the solar corona, which propagate through interplanetary space at velocities of $\sim$100--2500~km~s$^{-1}$. Although plane-of-sky coronagraph measurements have provided some insight into their kinematics near the Sun ($<$32~R$_\odot$), it is still unclear what forces govern their evolution during both their early acceleration and later propagation. Here, we use the dual perspectives of the Solar TErrestrial RElations Observatory (STEREO) spacecrafts to derive the three-dimensional kinematics of CMEs over a range of heliocentric distances ($\sim$2--250\,R$_{\odot}$). We find evidence for solar wind (SW) drag-forces acting in interplanetary space, with a fast CME decelerated and a slow CME accelerated towards typical SW velocities. We also find that the fast CME showed linear ($\delta=1$) dependence on the velocity difference between the CME and the SW, while the slow CME showed a quadratic ($\delta=2$) dependence. The differing forms of drag for the two CMEs indicate the forces and thus mechanism responsible for there acceleration may be different.
\end{abstract}


\keywords{Sun: coronal mass ejections (CMEs) --- Sun: activity --- solar-terrestrial relations}



\section{Introduction}
\label{Intro}


Massive eruptions of plasma and magnetic field which travel from the Sun through the Heliosphere are known as coronal mass ejections (CMEs). CMEs can have masses up to $10^{16}$\,g \citep{Vourlidas:2002p3006}, propagate at velocities of up to 2500\,km\,s$^{-1}$ \citep{Gopalswamy:2004p3007} close to the Sun, while at 1~AU velocities tend to be closer to that of the solar wind (SW; \citealt{Gopalswamy:2007p18}). Although CMEs have been the subject of study for nearly 40 years, a number of fundamental questions regarding their acceleration and propagation remain unanswered. One such question, what forces govern the propagation of CMEs in the Heliosphere has been especially difficult to tackle. This is mainly due to a lack of the three dimensional (3D) observations of CMEs in the inner Heliosphere.

The kinematic evolution of CMEs can be broken into three phases; initiation, acceleration, and propagation \citep{Zhang:2001p4793}. During the propagation phase, the initial acceleration has ceased and the CME motion is dominated by the interaction between the SW and the CME. The ``snow plough'', aerodynamic drag, and flux-rope models all aim to explain the motion of CMEs in the SW \citep{Tappin:2006p47,Borgazzi:2009p5016,Vrsnak:2010p7422,Cargill:2004p12,Chen:1996p5822}. An equation describing the motion of a CME in the drag dominated regime may be written:
\begin{equation}
\label{eq2}
M \frac{dv}{dt}= - 1/2 C_{D} \rho_{sw} A_{cme} (v-v_{sw})|v-v_{sw}|
\end{equation}
where $C_{D}$ is the drag coefficient, $\rho_{sw}$ is the solar wind density, $A_{cme}$ is the CME area and $v_{sw}$ is the solar wind velocity, and $M$ is the CME mass. We use a parametric drag model similar to that of \cite{Vrsnak:2002p3079} with the added parameter $\delta$, which determines if the drag is quadratic or linear. This parametric form collapses the complex dependences of the CME area ($A_{cme}$) and the solar wind density ($\rho_{sw}$) into a power-law which depends on heliospheric distance $R$. Eq. \ref{eq2} can thus be written
\begin{equation}
\label{eq3}
	\frac{dv}{dt}= - \alpha R^{-\beta}(v-v_{sw})^{\delta}
\end{equation}
where $\alpha$, $\beta$, and $\delta$ are constants.

Before the launch of the Solar TErrestrial RElations Observatory (STEREO; \citealt{Kaiser:2008p1663}) mission, synoptic white-light CME observations were limited to 32\,R$_{\odot}$ using Large Angle Spectrometric Coronagraph (LASCO; \citealt{Brueckner:1995p3300}), while the Solar Mass Ejection Imager (SMEI; \citealt{Jackson:2004p2417}, \citealt{Howard:2006p7460}) sometimes tracked CMEs to Earth ($\sim$215\,R$_{\odot}$). In radio observations, fast CMEs which drove shocks could be tracked to Earth \citep{Reiner:2007p5362}. Interplanetary Scintillation (IPS) observation provided density and velocity measurements for both CMEs and the SW from 50\,R$_{\odot}$ to beyond 1\,AU and using tomographic techniques can give 3D information \citep{Manoharan:2006p6629,Manoharan:2010p8035}. CMEs are also observed in {\it in-situ} measurements with WIND and ACE at L1 ($\sim$1\,AU), and occasionally CMEs can be tracked up to very large distances of up to 5\,AU using additional spacecraft \citep{Tappin:2006p47}. Numerical modelling has been used to study CME propagation with numerous approaches such as, 1D Hydro simulations, 2.5D MHD simulations and full 3D MHD simulations \citep{GonzalezEsparza:2003p8164,Cargill:1996p1291,Cargill:2002p1268,Odstrcil:1999p7707,Odstrcil:2004p8047,Smith:2009p8212,Falkenberg:2010p8153}.

Statistical studies comparing {\it in-situ} with white light observations indicate a trend of CME velocity converging towards the SW velocity as they propagate to 1\,AU \citep{Gopalswamy:2007p18}. Other studies, based on white light observations have indicated that aerodynamic drag of some form may explain this trend \citep{Vrsnak:2001p3953,Shanmugaraju:2009p5226}.  Radio observations suggest that a linear form of aerodynamic drag is most appropriate for fast CMEs \citep{Reiner:2003p5332}. \cite{Tappin:2006p47} showed that acceleration can continue far out (5\,AU) into the Heliosphere. However, these studies are subject to the difficulties associated with the observations they are based on. For example white light observations were limited to single, narrow, fixed, view-points meaning only observation of the inner Heliosphere could be made and even these were  subject to projection effects \citep{Howard:2008p15}. Also, linking features in imaging and {\it in-situ} observations is complex and can be ambiguous, a problem exacerbated during periods of high activity. In the case of numerical simulations their complexity can make it hard to extract which effects are the most important, possibly obscuring the important underlying physics.

The unique STEREO mission consists of two nearly-identical spacecraft in heliocentric orbits, STEREO-B(ehind) and STEREO-A(head) which separate from the Sun-Earth line at 22.5$^{\circ}$ per year. Each spacecraft carries the Sun Earth Connection Coronal and Heliospheric Investigation (SECCHI: \citealt{Howard:2008p4742}) suite, which images the inner Heliosphere from the Sun's surface to beyond 1~AU. Using STEREO observations, a number of papers have been published which extract 3D information and study CMEs at extended heliocentric distances, over-coming some of the difficulties outlined above. \citet{Davis:2009p6240} identified a CME in HI1 and HI2, using a constant velocity assumption \citep{Sheeley:2008p6253} they derived the speed and trajectory of the CME. The predicated arrival time, based on the speed derived, agreed with the {\it in-situ} observations. \citet{Wood:2009p4888} used the ``Point-P'' and ``Fixed-$\phi$'' methods to derive the height, speed, and direction from elongation measurements out to distances of $\sim$120\,R$_{\odot}$. A recent paper by \citet{Liu:2010p7366} tracked a CME to $\sim$150\,R$_{\odot}$ in 3D using J-maps from both spacecraft to triangulate the CMEs position in 3D. On the other hand \citet{Maloney:2009p6617} tracked the trajectory of CME apexes in 3D using triangulation, some as far as 240 R$_{\odot}$. \citet{Byrne:2010} developed a new reconstruction method with allowed the entire CME front to be reconstructed. They found evidence for CME deflection, expansion and acceleration low down ($<$\,7\,R$_{\odot}$) followed by a solar wind drag interaction. For a review of some of the different 3D reconstruction methods which have been applied to STEREO CME observations see \citet{Mierla:2010p7463}.

In this paper, we use triangulation to localise CME apexes in 3D. From this, we derive the CME apex trajectory and kinematics. These kinematics are then used to investigate the effects of drag on the CME.  We present the reconstructed CME (apex) kinematics for three events, one acceleration, one decelerating, and one with constant velocity. In Section 2 we describe the observations, data reduction, and the reconstruction and fitting technique. Section 3 includes a discussion of each event in detail and presents the reconstructed kinematics themselves. The implications of our results and our final conclusions are given in Section 4.

\section{Observations and Data Analysis}
\subsection{Observations}
The trajectories of three CMEs were reconstructed using observations from STEREO SECCHI. SECCHI consists of five telescopes, the Extreme Ultraviolet Imager (EUVI), the inner and outer coronagraphs (COR1 and COR2), and finally the Heliosphereic Imager (HI1 and HI2). COR1 images the corona from 1.4--4.0~R$_{\odot}$, while COR2 images the corona from 2.5--15~R$_{\odot}$. Both of the coronagraphs take sequences of  three polarised images which can be combined to give total brightness (B) or polarised brightness (pB) images \citep{Howard:2008p4742,Thompson:2003p1587}. The HI instrument is a combination of two refractive optical telescopes with multi-vein, multi-stage light rejection system which images the inner Heliosphere from 4--89 degrees \citep{Eyles:2008p3861}. HI1 images the inner Heliosphere from 3.98--23.98$^{\circ}$ (degrees elongation) in white light with a cadence of 40 minutes while HI2 images the Heliosphere from 18.68--88.68$^{\circ}$ in white light with a cadence of 2 hours.

The three CMEs considered here were observed during: 2007 October 8--13 (CME 1), 2008 March 25--27 (CME 2), and 2008 April 9--12 (CME 3). The observations were reduced using {\it secchi\_prep} from the {\sc SolarSoft} library \citep{Freeland:1998p3546}.  This consisted of debasing and flat-fielding for all images. The COR1 and COR2 images were also corrected for vignetting, exposure time, and an optical distortion. The COR1 observations had a model background subtracted to remove static coronal features. The HI instrument has no shutter, and as such, these observations needed additional corrections for smearing and pixel bleeding. The pointing of the HI observations were updated using known star positions within the filed-of-view \citep{Brown:2008p3595}. Standard running difference images were created from the COR1/2 observations while a specialised running difference technique was used to suppress the stars for the HI observations \citep{Maloney:2009p6617}.  The relative drift, due to satellite motion, of the star field between two successive HI images is calculated and then the earlier image is shifted to account for this motion removing a large part of the background signal. \figref{1} shows reduced observations from the 2008 March 28 event where the CME is simultaneously observed in both COR1 and COR2 in both the Ahead and Behind spacecraft but only in from the Ahead spacecraft in HI.

\begin{figure*}[ht]
\label{fig1}
\includegraphics[trim = 0 20 0 450, clip, scale=0.90]{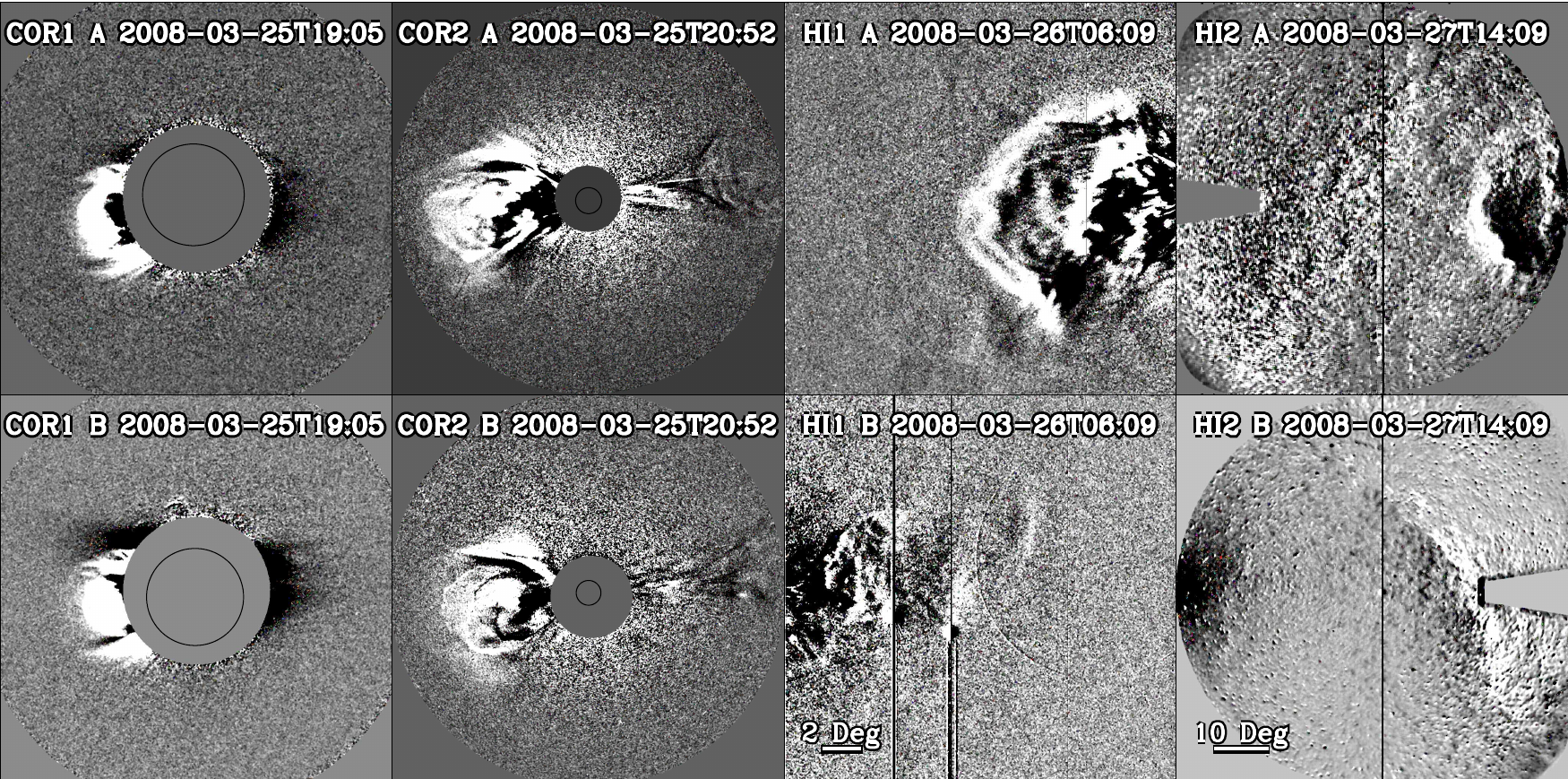}
\caption{Sample images  of the 2008 March 25 event showing the CME in COR1, COR2, HI1, and HI2 fields-of-view. Observed from STEREO-A (top row) and from STEREO-B (bottom row). Note the absence of a clear CME signature in the STEREO-B HI1 and HI2 images.}
\end{figure*}

\subsection{3D Reconstruction}
Each event was observed in either the inner coronagraph (COR1) or outer coronagraph (COR2) simultaneously by both STEREO-A and STEREO-B. From these images, the CME apex was localised via tie-pointing (see \citealt{Maloney:2009p6617} Figure 1, \citet{Inhester:2006p2249}). The trajectory was then reconstructed by tracking it through a series of images. In all the events presented, the CME was only observed in HI by one spacecraft, so an additional constraint was required to localise the CME apex. We therefore assumed that the CME continued along the same path with respect to solar longitude, as it did in the the COR1/2 field-of-view (i.e. travelled radially; \citealt{Maloney:2009p6617}). Figure 2 shows the derived trajectory for the 2008 March 25 event. Once the 3D trajectories were derived, we calculated the height  and then took numerical derivatives with respect to time to obtain the velocity and acceleration.

\begin{figure}[ht!]
\label{fig2}
\includegraphics[trim = 10 20 80 180, clip, scale=0.5]{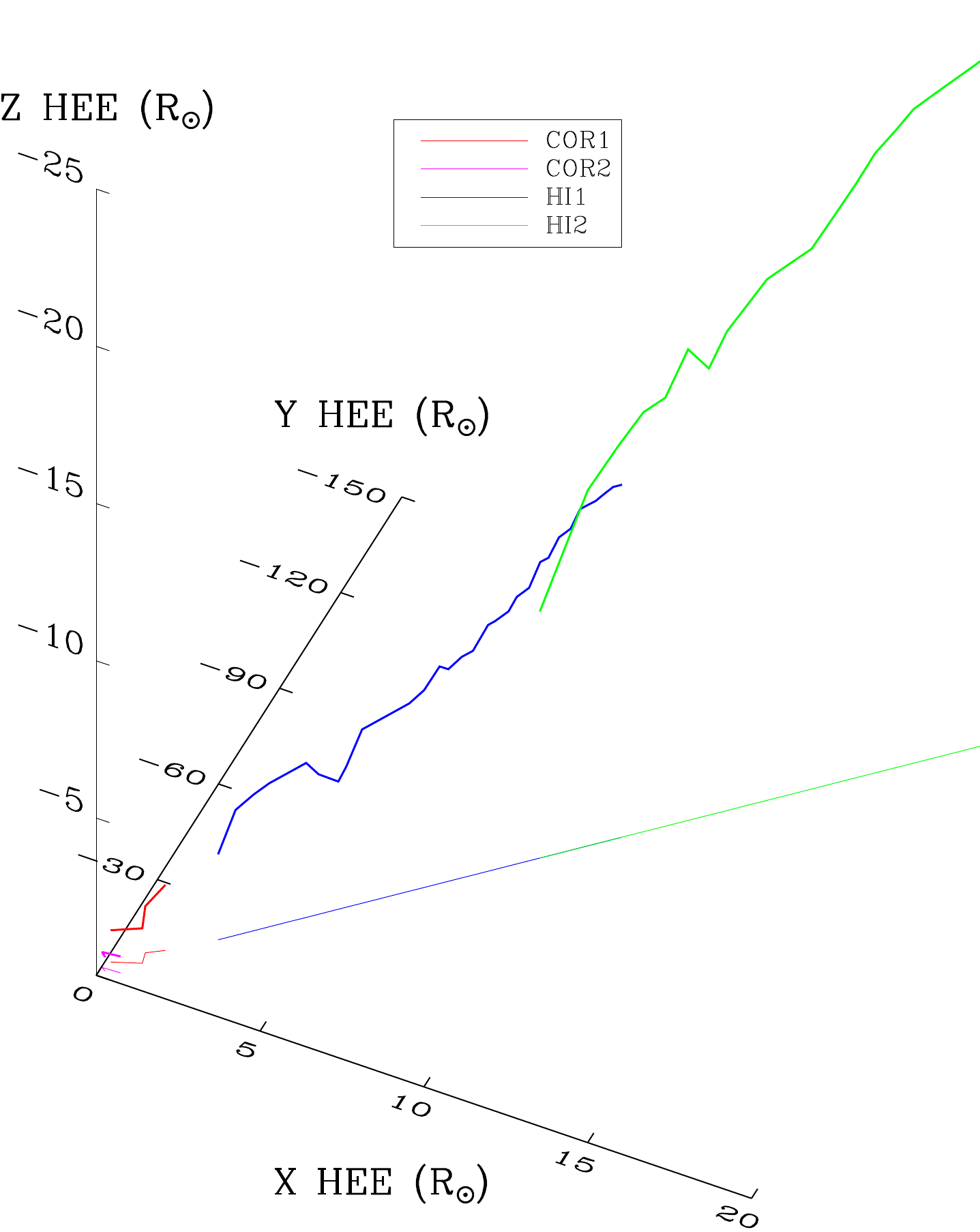} 
\caption{Reconstructed apex trajectory of the 2008 March 25 CME. The CME is tracked through COR1, COR2, HI1 and HI2 fields-of-view. For the COR1 and COR2 observations the CMEs position in reconstructed using triangulation. In the HI field of view we assume the CME will continue along the same path as it did in COR1 and COR2 (traces in x--y plane) in order to determine the CMEs position in 3D space. The x-axis points towards the Earth, the y-axis is is perpendicular to x-axis and in the ecliptic plane, the z-axis is perpendicular to both.}
\end{figure}

\subsection{Kinematic Modeling}
The kinematics were fitted, via a least-squares method, with a parametric model for the drag (Eq. \ref{eq3}). In order to test which form of drag is most suitable (linear or quadratic), we fitted (Eq. \ref{eq3}) with $\delta$ set to 1 and then separately with $\delta$ equal to 2. The kinematics were only fit during the time interval we believe that drag is at play and the observations are accurate. There was evidence for an early acceleration phase not attributed to drag which was not fitted.  Also, events which were tracked far into HI2 field-of-view where identification of the CME apex become ambiguous were excluded from fitting.

A number of the model parameters can be fixed from the observations, such as the CME height and velocity. We assume that the CME tends to the SW speed, which was taken to be where which the velocity plateaus. The model parameters obtained from the fitting were then compared with previous results from \cite{Vrsnak:2002p3079}. From this comparison, we infer which model best reproduces the kinematics and hence is the most appropriate. Both the fast and slow CMEs (CME 1 and CME 2) were analysed using this method. The intermediate CME (CME 3) was fitted with a constant acceleration model to show that there was no significant acceleration involved.

\section{Results}
\label{Res}
CME 1 and CME 2 were fit in three ways with $\alpha$, $\beta$, and $\delta$ all allowed to vary (black line), with fixed $\delta$ of two (magenta line) and one (orange line). In both cases the free fitting returned values that were not comparable to previous studies \citep{Vrsnak:2001p3953}. The different fit parameters for the events are given in Table 1. CME 3 was fit with a constant acceleration model (black line). 

\subsection{CME 1 (2007 October 8--13)}
 \figref{3}(a)-(c) shows the kinematics for the accelerating CME. This CME was first observed at 15:05~UT on 2007 October 8 off the west limb and was found to be propagating at an angle of 56$^{\circ}$ from the Sun-Earth line. \figref{3}(a) shows the height of the CME. \figref{3}(b) shows the velocity profile which clearly shows the CME is undergoing acceleration, initial velocity of $\sim$150~km~s$^{-1}$ and final velocity of $\sim$450~km~s$^{-1}$. There may be two acceleration regimes, an early increased acceleration phase (before 18:00~UT on the October 8) followed by a drag acceleration. The early acceleration can be attributed to a magnetic driving force and so was not fitted with the drag model. Later, when the CME reached the centre of the HI2 field-of-view, determining the front position becomes difficult so this region was not fitted. \figref{3}(c) shows the acceleration profile of the event. The $\delta=2$ (orange) fit givies the lowest chi-squared value.

\begin{figure*}
\label{fig3}
\caption{{\it Left panel:} Kinematics for CME 1, (a) height, (b) velocity, and (c) acceleration. {\it Right panel:} Kinematics for CME 2, (d) height, (e) velocity, and (f) acceleration. Vertical dashed line indicates separation between early and late phase acceleration. Horizontal dot-dash line indicates the inferred SW velocity.}
\includegraphics[scale=0.4]{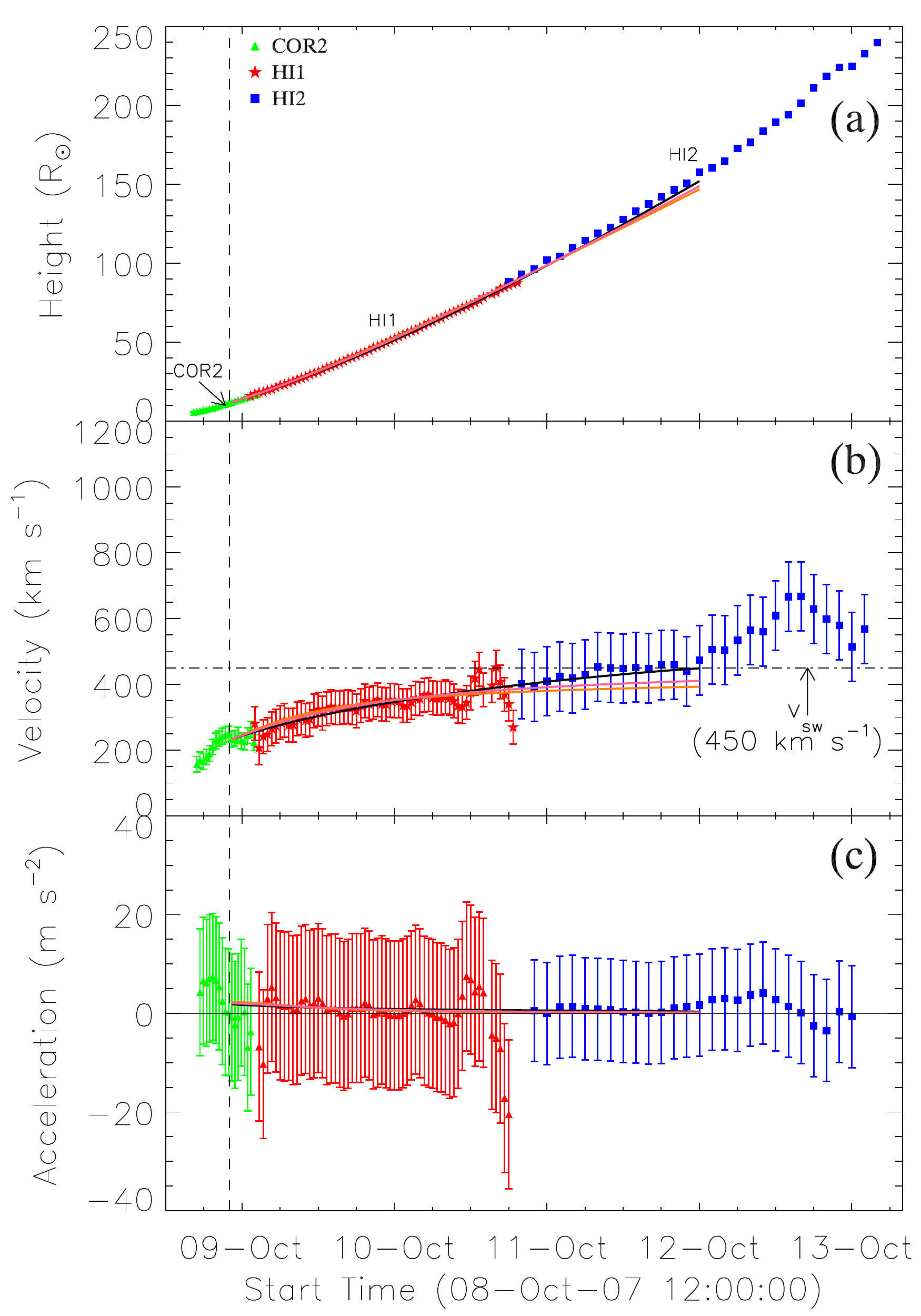}
\includegraphics[scale=0.4]{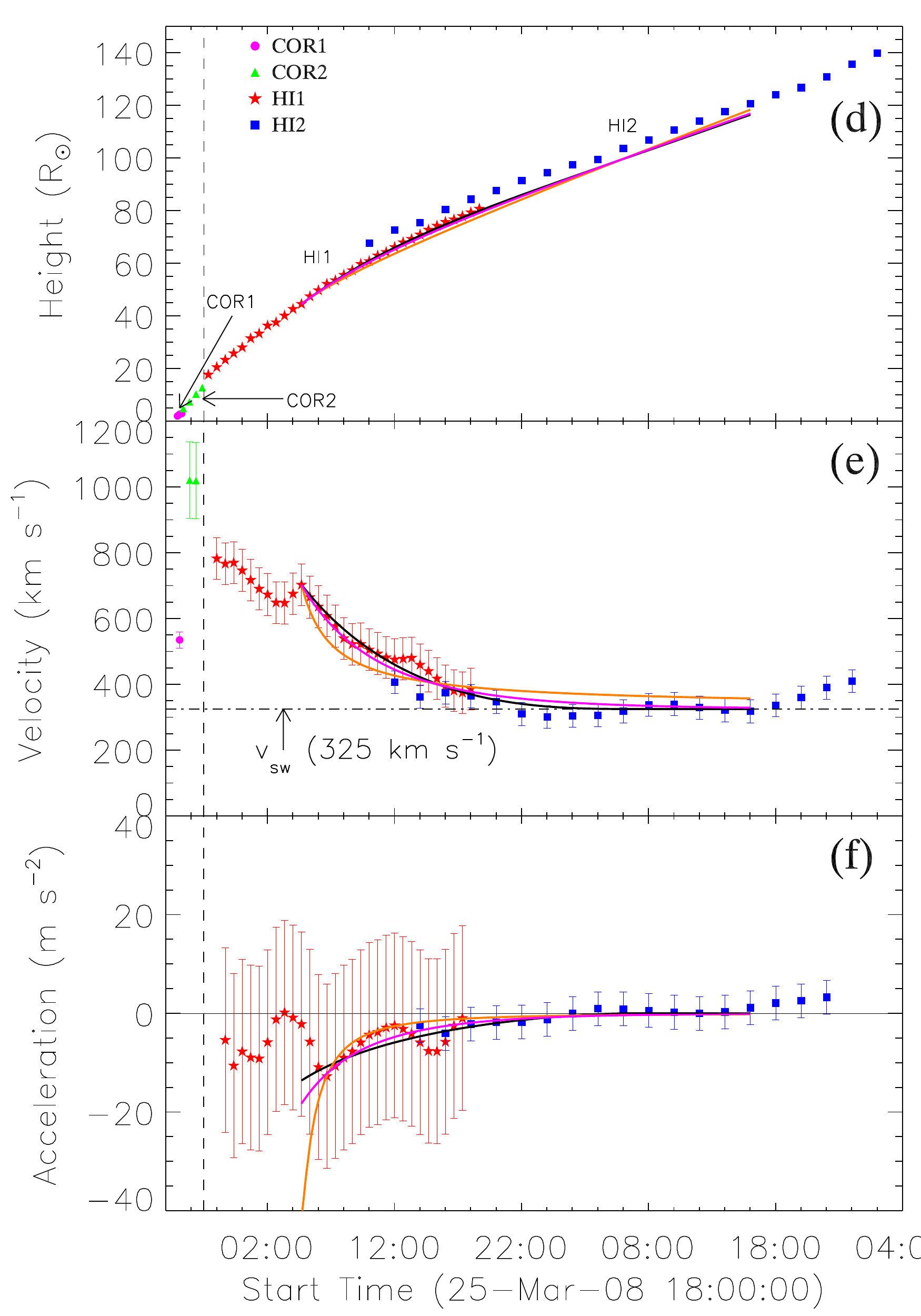}
\end{figure*}

\subsection{CME 2 (2008 March 25--27)}
The kinematics from the decelerating CME are shown in \figref{3}(d)-(f). This CME was first observed at 18:55~UT on 2008 March 25 off the east limb and was found to be propagating at an angle of -82$^{\circ}$ from the Sun-Earth line. \figref{3}(d) and (e) show the height and velocity profiles, the velocity profile clearly demonstrates the CME is undergoing deceleration. The CME had an initial velocity (HI1) $\sim$\,800\,km\,s$^{-1}$ and final velocity of $\sim$\,375\,km\,s$^{-1}$. Due to the high speed of this CME, it was only observed in a small number of frames in COR1 and COR2. As a result, the kinematics were difficult to quantify in these instruments. However, there appears to have been an early acceleration feature. The deceleration in the HI1 and HI2 field-of-view continued until the CME reaches a near-constant velocity, and travels at this velocity throughout the rest of the field-of-view. \figref{3}(d) shows the acceleration profile of the event. The $\delta=1$ (magenta) fit gives the lowest chi-squared value.

\subsection{CME 3 (2008 April 9--12)}
In \figref{4} we show the kinematics of the constant velocity CME. This CME was first observed at 15:05~UT on 2008 April 09 off the east limb and was found to be propagating at an angle of -73$^{\circ}$ from the Sun-Earth line. \figref{4}(a) shows the height of the CME \figref{4}(b) shows the velocity profile which has a scatter about $\sim$\,300\,km\,s$^{-1}$. Again, there may be some evidence in the COR1/2 observations for an early acceleration phase but due the events poorly observable features, at this early stage, it is hard to quantify this. The departure from the fit after April 12 20:00~UT is thought to be due to error in the reconstruction as the CME apex becomes to faint to identify.  As this event shows no obvious acceleration it was not fitted with the drag model but with a constant acceleration model $h(t) = h_{0}+ v_{0}t + 1/2 at^2$ (thin black line). \figref{4}(c) shows the acceleration profile, the fit values ($h_{0}$\,=\,22\,R$_{Sun}$,  $v_{0}$\,=\,334\,km\,s$^{-1}$, and a\,=\,$-0.18$\,m\,s$^{-2}$) are consistent with no acceleration throughout the field-of-view.
\begin{figure}[h!]
\label{fig4}
\caption{Kinematics for CME 3 (a) height (b) velocity (c) acceleration. This event shows an early acceleration (to left if dashed line) which levels of to a scatter about typical solar wind speeds. This event was fit with a constant acceleration the resulting fit parameters are $h_{0}$\,=\,22\,R$_{Sun}$,  $v_{0}$\,=\,334\,km\,s$^{-1}$, and a\,=\,$-0.18$\,m\,s$^{-2}$. The assumed solar wind value is indicated by the horizontal dot-dash line}
\includegraphics[scale=0.48]{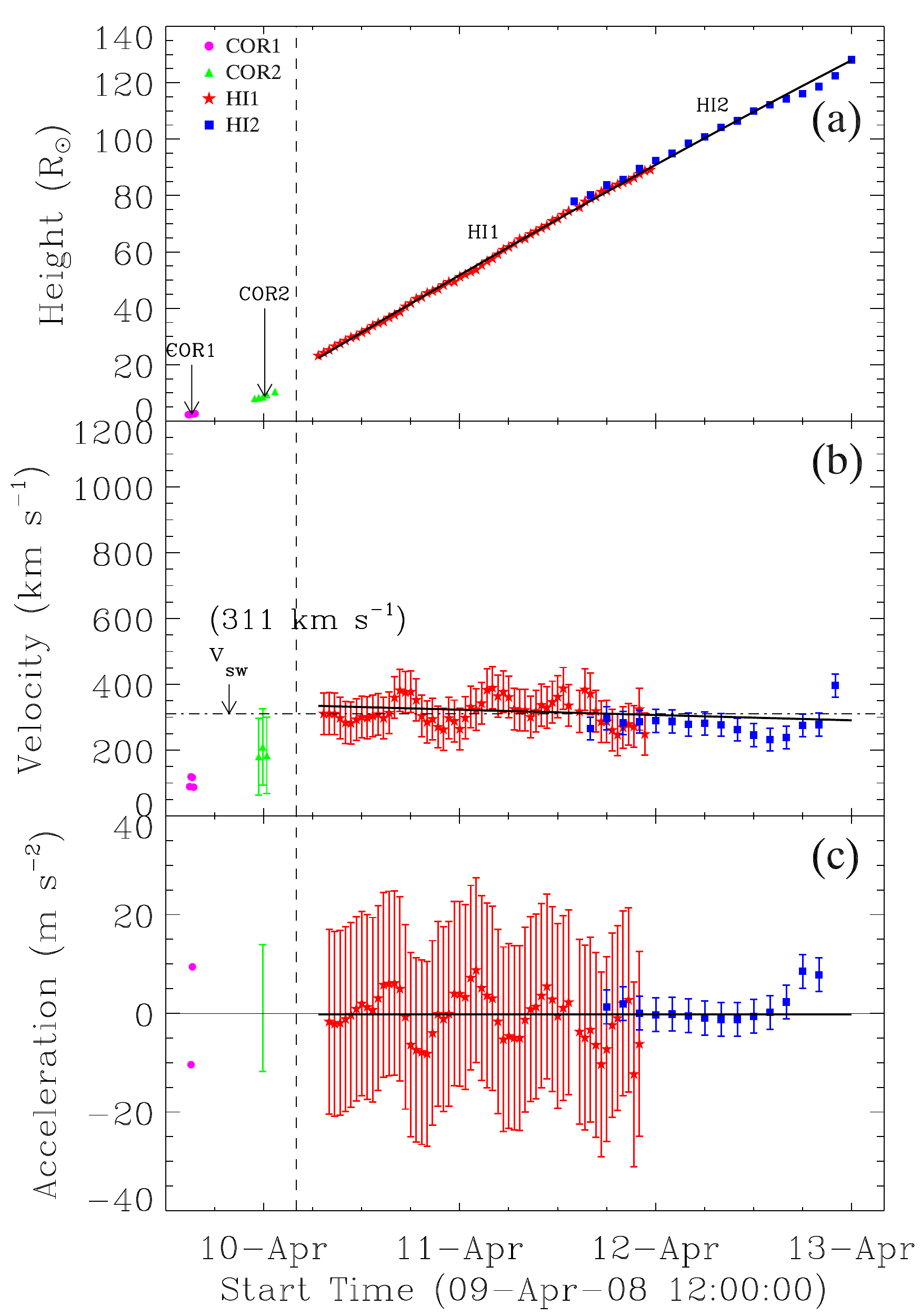}
\end{figure}

\begin{deluxetable}{lcccc}
	\tabletypesize{\scriptsize}
	\tablewidth{0pt}
	\tablecaption{Fit parameters for the accelerating events. CME 1, $v_{sw}=450$\,km\,s$^{-1}$, $v_{cme}=233$\,km\,s$^{-1}$ , $h_{0}=12$\,R$_{\odot}$. CME 2, $v_{sw}=325$\,km\,s$^{-1}$, $v_{cme}=702$\,km\,s$^{-1}$ , $h_{0}=44$\,R		$_{\odot}$.}
	\tablehead{\colhead{} & \colhead{$\alpha$} & \colhead{$\beta$} & \colhead{$\delta$} &  \colhead{$\chi^{2}$} \\} 

	\startdata
		CME 1 (2007  0ct  8)  \\
		\phantom{aaa}Linear (magenta) & 1.61e-5 & -0.5 & 1.0 & 8.27  \\
		\phantom{aaa}Quadratic (orange) & 1.28e-7 & -0.5 & 2.0 & 6.74 \\
		\\
		CME 2 (2008  Mar  25) \\
		\phantom{aaa}Linear (magenta) & 1.02e-4 & -0.5 & 1.0 & 3.71 \\
		\phantom{aaa}Quadratic (orange) & 6.38e-7 & -0.5 & 2.0 & 17.63 \\
	\enddata
\end{deluxetable}

\section{Discussion and Conclusions}
\label{DisConc}
We have shown it is possible to derive the 3D kinematics of features, the CME apex in this case, in the inner Heliosphere ($\sim$2--250\,R$_{\odot}$) using STEREO observations. The 3D kinematics are free from the projection effects of traditional 2D kinematics but may contain artifact from the 3D reconstruction method (e.g, \citealt{Maloney:2009p6617}) and other sources. Both of the accelerating events showed two regimes in the velocity profile, a low down ($<$\,15\,R$_{\odot}$) early rapid acceleration (in comparison to later values), followed by a gradual acceleration far from the Sun ($>$\,30\,R$_{\odot}$). The early acceleration is thought to be due to a magnetic driving force, as the solar wind velocity low in the corona $(v_{sw}(\le10$\,R$_{\odot})\le$\,268\,km\,s$^{-1}$, \citealt{Sheeley:1997p3966}) is lower than the velocity already attained by the CMEs in both cases. Here we assume that the later acceleration is due the interaction between the SW and the CME, as in each case the CME attains a final velocity close to typical values for the solar wind.

Considering CME 2 in \figref{3}(d)-(f), it can clearly be seen that the velocity levels off to a constant value typical of the solar wind. We interpret this as the CME reaching the local solar wind speed, as a result the force acting on the CME goes to zero. For CME 1 \figref{3}(a)-(c), the velocity initially increases, however, there is a plateau towards the end after April 11 6:00 UT which occurs at SW like speeds. The height measurements towards the end is very scattered and shows rapid increase. This is most likely due to losing the front to the background noise and triangulating a different feature. CME 3 propagates at a roughly constant velocity, which is consistent with the drag interpretation. The CME appears to have already attained the local SW speed and therefore is not accelerated. The fitting results show that a linear dependence produces a better fit for the fast event (CME 2), while a quadratic dependence better fits the slow event (CME 1). The differing range of the interaction CME 1 $\sim$\,120\,R$_{\odot}$ and CME 2$\sim$ 80\,R$_{\odot}$ may be explained by the suggestion that wide low mass CMEs are more affected by drag than narrow massive CMEs \citep{Vrsnak:2010p7422}.

\cite{Reiner:2003p5332} suggest that for fast events, a linear model of drag better reproduces the kinematics, which agrees with our findings. \cite{Vrsnak:2001p3953} also suggested that a linear dependence might be appropriate, however the quadratic form has been studied much more. From a theoretical perspective, a quadratic dependence corresponds to aerodynamic drag, while a linear dependence suggests Stokes' or creeping drag. It is not currently clear which model is more physically correct. The fit parameters obtained do not agree with those found by \cite{Vrsnak:2001p3953} and while our values are not unphysical, it is not clear why they differ so much from the previous studies.

The mechanism behind the apparent differing forms of drag, linear ($\delta=1$) and quadratic ($\delta=2$), for the slow and fast event are unclear. The application of any hydrodynamic theory to a CME, such as drag, may be missing vital physics. Could the magnetic properties play a role modifying the form of the drag (reconnection, suppression of turbulence, wave energy transport)? For example \cite{Cargill:1996p1291} showed that depending on the orientation of the flux rope and background magnetic field (aligned or non-aligned) the drag coefficient can vary between zero and ~3. They also found that the magnetic field of the flux rope is important in order for its survival as it propagates. Further which form of drag is correct for a CME in the SW, the low Reynolds number viscous dominated Stokes'  drag or the high Reynolds number turbulence dominated  aerodynamic drag? In order to address these questions a larger sample study is needed in order to verify these effects are recurring and observable phenomena and also to build up the statistics.

We have shown it is possible to derive the true 3D kinematics for a number of CMEs in the inner Heliosphere. Based on this we have been able to conclusively show that CMEs undergo acceleration in the inner Heliosphere, more specifically, that due to its range and strength this acceleration is believed to be the result of some form of drag. This drag acceleration has important implications for space weather predictions and for the analysis techniques which assume CMEs travel at constant velocity through the Heliosphere. The HI observations of CMEs in the Heliosphere provide a unique and limited opportunity to study the propagation of CMEs and to understand the coupling between the solar wind and CMEs.






\acknowledgments

This work is supported by Science Foundation Ireland Grant No. 07-RFP-PHYF399.
We would also like to thanks the STEREO/SECCHI consortium for providing open access
to their data and technical support.

\clearpage



\clearpage



\end{document}